\documentclass{aa}
\usepackage{graphicx}
\usepackage{natbib}
\newcommand{\map}{M_{\mathrm{ap}}}
\newcommand{\be}{\begin{equation}}
\newcommand{\ee}{\end{equation}}

\newcommand{\myarcsec}{\hbox{$.\!\!^{\prime\prime}$}}
\newcommand{\myarcmin}{\hbox{$.\!\!^{\prime}$}}

\newcommand{\myarcminnodot}{\hbox{$\;\!\!^{\prime}\;$}}
\renewcommand{\l}{\left}
\renewcommand{\r}{\right}
\begin{document}
   \title{GaBoDS: The Garching-Bonn Deep Survey}

   \subtitle{II. Confirmation of EIS cluster candidates by weak
     gravitational lensing \thanks{Based on observations made with ESO
     Telescopes at the La Silla Observatory}}
   \author{Mischa Schirmer\inst{1}, Thomas Erben\inst{1}, Peter
     Schneider\inst{1}, Christian Wolf\inst{2}, Klaus Meisenheimer\inst{3} 
          }

   \offprints{M. Schirmer}

   \institute{Institut f\"ur Astrophysik und Extraterrestrische 
              Forschung (IAEF), Universit\"at Bonn, Auf dem H\"ugel 71, 
              53121 Bonn, Germany; \email{mischa@astro.uni-bonn.de}
	 \and
              Department of Physics, Denys Wilkinson Bldg., University of
              Oxford, Keble Road, Oxford
	 \and
             Max-Planck-Institut f\"ur Astronomie, K\"onigstuhl 17, 69117
              Heidelberg, Germany
             }

   \date{Received ??-??-2004; accepted ??-??-2004}

   \abstract{We report the first confirmation of colour-selected galaxy 
   cluster candidates by means of weak gravitational lensing. Significant 
   lensing signals were identified in the course of the shear-selection 
   programme of dark matter haloes in the Garching-Bonn Deep Survey, which 
   currently covers 20 square degrees of deep, high-quality imaging data on 
   the southern sky. The detection was made in a field that was previously
   covered by the ESO Imaging Survey (EIS) in 1997. A highly significant 
   shear-selected mass-concentration perfectly coincides with the richest EIS 
   cluster candidate at $z\approx0.2$, thus confirming its cluster nature. 
   Several other shear patterns in the field can also be identified with 
   cluster candidates, one of which could possibly be part of a filament at 
   $z\approx0.45$.
   }

\titlerunning{Confirmation of cluster candidates by weak lensing}
\authorrunning{M. Schirmer et al.}

   \maketitle

\section{Introduction}
The search for dark matter haloes by weak gravitational lensing was first 
proposed by \cite{zeta} and \cite{schneider_map} (hereafter S96). Using 
this technique, clusters are detected directly by their mass instead of 
their luminosity, and no assumptions have to be made about their virial state
or the relation between bright and dark matter. In this way, a mass-selected 
cluster sample can be established, and the cluster population can be probed
e.g. for underluminous objects. With moderately sized telescopes, however,
this method is restricted to haloes in the intermediate redshift range (up to 
$z\approx0.6$), since with an increasing lens distance the population of 
sheared background galaxies is shifted to higher redshifts, making their shape 
measurement increasingly difficult.

\cite{kruse} predicted the number density of shear-selectable galaxy clusters 
with $S/N\geq5$ to be of the order of 10 deg$^{-2}$. This figure is based on 
optimistic estimates of the width of the ellipticity distribution of galaxies 
and their accessible number density. With 2m-class telescopes, the number of 
such clusters per deg$^{-2}$ is about a factor of 5-10 smaller. Since the
advent of wide field imagers, a number of shear-selected dark matter haloes
were reported in the literature \citep[see][for example]
{erben_darkclump,umf00,maoli,wtm01,wtm02,meh02,miyazaki,dahle,schirmer03}.
About half of these objects have optical counterparts, the nature of the
others remains unclear.

In this work we present a sample of shear-selected dark matter haloes which
(mostly) coincide with colour-selected galaxy cluster \textit{candidates}. 
The objects were found in one of the 63 fields of the Garching-Bonn Deep 
Survey (hereafter GaBoDS), which are currently analysed in this respect. 
Section 2 summarises the observations and the data reduction, and Sect. 3 
presents the \textit{mass aperture statistics} (see S96) used for the shear 
selection. The particular detections are presented in Sect. 4, and we draw
conclusions in Sect. 5.

\section{Observations and data reduction}
GaBoDS is a mostly virtual survey, with 80\% of the data taken from the ESO 
archive or contributed by collaborators. It covers a sky area of currently
20 square degrees taken with the Wide Field Imager (WFI) at the MPG/ESO 2.2m 
telescope. The outstanding image quality of this instrument makes it ideally 
suited for observations of the weak lensing effect.

The field (``SGP'') under consideration is centred on the South Galactic Pole
and was observed during MPG time in the course of the COMBO-17 project 
\citep[see][]{combo17}. Observations in $R$-band were carried out in October
1999 and September 2000. The total exposure time amounts to 27 ksec (7.5 
hours) dark time, 20 ksec of which were selected for the coaddition due to 
their sub-arcsecond seeing quality and PSF properties.

\setcounter{topnumber}{1}
\begin{figure}[t]
  \includegraphics[width=0.9\hsize]{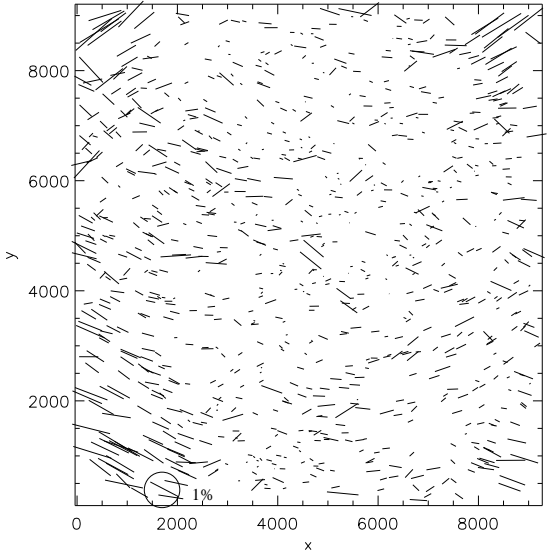}
  \caption{\label{SGP_aniso}PSF anisotropies in the stacked SGP image. The
  pattern is mostly due to slight misalignments of the CCDs with respect to 
  the focal plane.}
\end{figure}

The data reduction was performed with the GaBoDS pipeline, which was
specifically developed for the reduction of any kind of WFI data. A detailed 
description of the algorithms involved and their performance can be found in 
\cite{schirmer03}. Specific care was taken for the relative astrometric 
calibration of the individual CCDs. The accuracy obtained is on the order of a 
$1/10$th-$1/20$th of a pixel, thus no artificial anisotropies are introduced 
into the PSF (Fig. \ref{SGP_aniso}) that could mimic a shear signal. The
seeing in the coadded image (35\myarcminnodot$\times$ 37\myarcminnodot) is
0\myarcsec8, and we reach a limiting magnitude of $R=26.1$ (Vega) for objects
with at least 4 connected pixels $2.5\,\sigma$ above the sky background. The
measurement of the galaxy shapes and the correction for PSF effects was
performed with the KSB software \citep{ksb}. A detailed description and test
of our implemented shape measurement approach is given in \cite{howaccurate}. 

\section{The shear-selection method}
In the following, standard weak lensing notations are used. For a technical
review of this topic see \citet{bs_review}. 

The tidal gravitational field of a cluster-sized mass concentration induces a
coherent distortion pattern in the images of distant background galaxies. By
scanning the field for such characteristic patterns, the causing mass 
concentrations can be found. For the detection of these shear patterns we use
the \textit{mass aperture statistics} ($\map$, see S96). There, $\map$ is
defined as a filtered integral of the projected lens mass distribution
$\kappa$ inside an aperture. Switching to a different filter function,
$\kappa$ can be replaced by the tangential shear $\gamma_{\rm t}$, for which
the observable ellipticities $\epsilon_{\rm t}$ of the background galaxies are
an unbiased estimator. On a discrete grid $\map$ evaluates as
\begin{equation}
\map=\frac{\sum_i \varepsilon_{{\rm t}i}\,w_i\,Q_i}{\sum_i \,w_i}\,,
\label{map_def}
\end{equation}
where $\varepsilon_{{\rm t}i}$ are the tangential components of the 
ellipticities of the lensed galaxies. The $w_i$ are individual weighting
factors as introduced by \cite{howaccurate}, and $Q_i$ is the $\map$ filter
function. The noise for $\map$ is obtained as
\begin{equation}
{\sigma^2(\map)} = \frac{\sum_i |\epsilon_i|^2\, {w_i}^2\,{Q_i}^2}
                   {2\,\l(\sum_{i}\,w_i\r)^2}\,,
\label{sn_analytic}
\end{equation}
so that the ratio $S=\map/\sigma(\map)$ directly estimates the signal-to-noise
of the shear-selected dark matter halo detection. We call this quantity the 
\textit{S-statistics}, and its two-dimensional graphical representation the
\textit{S-map}.

\begin{figure}[t]
  \includegraphics[width=0.9\hsize]{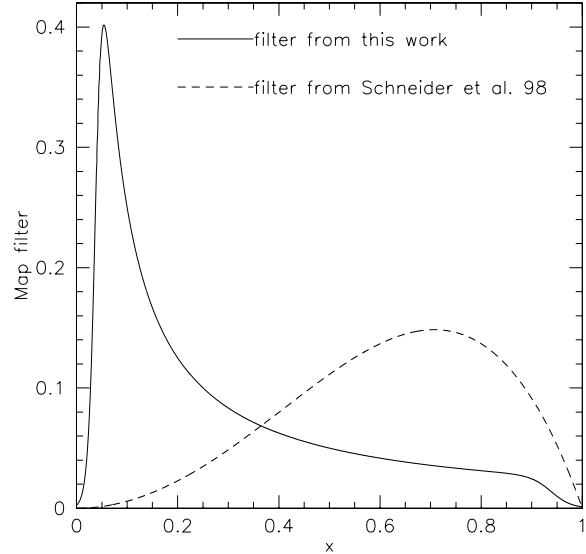}
  \caption{\label{MAP_filter}The plot compares the $\map$ filter
from this work with that proposed in \cite{schneider1998}. The x-axis
shows the distance from the aperture centre in units of the aperture size. 
Our filter closely resembles the expected tangential shear from a NFW profile, 
hence optimising the $S/N$ for such haloes. It is cut-off for very small and
very large radii.}
\end{figure}

We use a different filter function $Q$ than the ones proposed in S96 and in
\cite{schneider1998}, approximating the expected tangential shear profile of a 
NFW halo (see Fig. \ref{MAP_filter}).

Note that $\map$ is not any more directly related to the mass inside the
aperture once it is evaluated near the border of the field, or on a galaxy 
distribution with swiss-cheese topology (e.g. due to masking of bright
stars). However, the $S$-statistics itself is unaffected by this and can be
used to detect dark matter haloes, since $\map$ still estimates the amount of
shear alignment inside the aperture. More details of this and a thorough
discussion of filter functions will be found in Schirmer et al., 2004 (in
prep.), in which a shear-selected sample of about 100 mass concentrations 
based on the 20 square degrees of GaBoDS will be presented.
\begin{figure}
  \includegraphics[width=1.0\hsize]{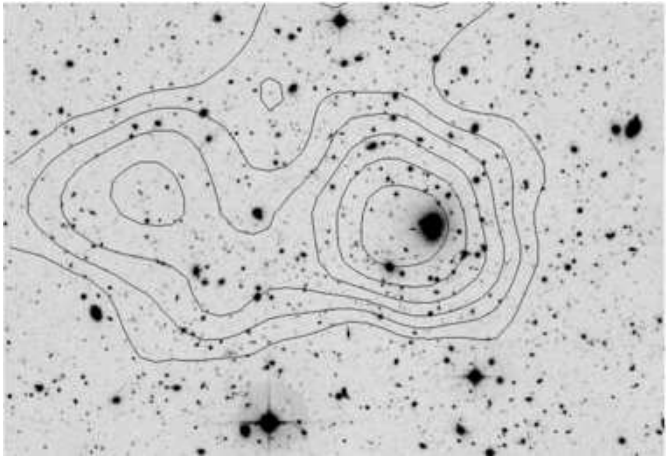}
  \caption{\label{SGP_peaks_2}Shear-selection of EIS0045$-$2923, detected at
  the $5.0\sigma$-level. The significance contours start with $2\sigma$
  and increase in steps of $0.5\sigma$. The field shown is 7\myarcmin5 wide.
  The full field is shown in Fig. \ref{SGP_peaks_1}. One needs to keep in mind
  that these contours do not show a mass reconstruction, but the significance
  of the tangential shear found around the position under consideration.}
\end{figure}

\section{Results}
\begin{figure*}
  \center\includegraphics[width=0.82\hsize]{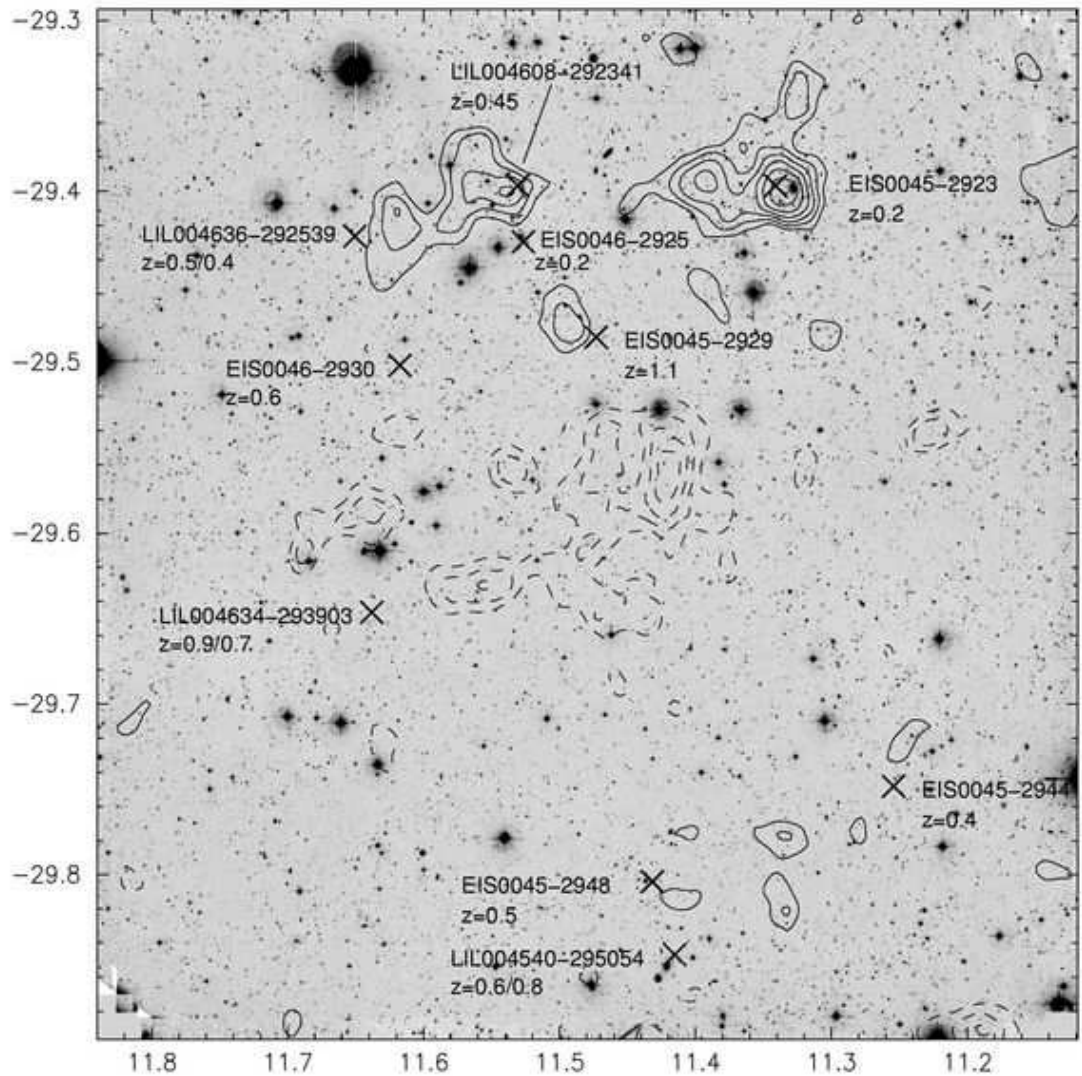}
  \caption{\label{SGP_peaks_1}$S$-statistics of the SGP field. The 
  contours start at the $2\sigma$-level and go in steps of $0.5\sigma$. 
  Dashed lines denote negative signals, i.e. underdense regions. The extended 
  depression in the centre of the field is significant up to the 
  $4\sigma$-level. The colour-selected cluster candidates are indicated by 
  crosses. The axes denote RA and DEC in degrees. North is up and East is
  left.}
\end{figure*}

Our weak shear detections in the SGP field are presented in Fig. 
\ref{SGP_peaks_1}. Shown are the contours of the $S$-map for a filter 
scale of 20\myarcminnodot . The number density of 
galaxies was $n=24$ arcmin$^{-2}$. The optical counterparts were selected in 
$V$ and $I$ by \cite{olsen} (hereafter EIS) and \cite{lil} (hereafter LIL), 
both using the EIS Wide Survey Patch B field and the matched filter algorithm 
by \cite{postman}. The authors estimate the redshifts of these candidates 
based on the colour of their elliptical galaxies.

Most prominently, EIS0045$-$2923 ($z\approx0.2$) is detected at the 
$5.0\sigma$-level (Fig. \ref{SGP_peaks_2}). This is also the richest cluster 
candidate as identified by EIS, even though it looks rather poor in the 
$R$-band image (as do all the other candidates), with a single cD galaxy. From 
its estimated redshift and the lensing $S/N$ we conclude that this object has 
a mass of a few $10^{14}$ ${\rm M_\odot}$. A significant extension to the 
east (left) is visible, in the middle of which about half a dozen ellipticals 
at possibly higher redshift are found.

11\myarcminnodot east of EIS0045$-$2923, a $3.6\sigma$-detection coincides 
with LIL004608$-$292341 ($z\approx0.45$). The galaxies therein form two 
strongly interacting subgroups. 1\myarcmin5 further to the South is
EIS0046$-$2925 ($z\approx0.2$), which is not seen in any of our $S$-maps. The 
lensing contours extend towards the East on a significant level for filter 
scales from 7-20 arcmin. 6\myarcmin5 (2.2 Mpc) into this direction we have 
LIL004636$-$293539 ($z\approx0.4-0.5$), falling slightly outside the 
$S$-contours. If these two clusters are indeed at the same redshift, this 
could indicate the presence of a filament. For another weak lensing detected 
filament see for example \cite{gray}.

When looking at the $S$-map for the full SGP field, the detections for 
EIS0045$-$2923 and LIL004608$-$292341 stand out remarkably clear over the rest
of the field, as well as in their size as in their significance. In Fig. 
\ref{SGP_peaks_1} several other EIS and LIL cluster candidates are indicated. 
The $2\sigma$-detection next to EIS0045$-$2948 in Fig. \ref{SGP_peaks_1} 
rises up to $4.3\sigma$ for a smaller filter scale of 6\myarcmin3, but is 
offset by 1\myarcminnodot to the West. This is consistent with the rms offset
we find for other mass concentration/optical counterpart pairs in our survey. 

In order to rule out any remaining systematic effects inherent to the data,
we split the 68 raw WFI images into two halves. One half contained only images 
with excellent PSF properties (better than 3\% anisotropy), whereas the second 
one consisted of those with anisotropies up to 6\% (mostly due to a 
slight defocusing of the telescope). The PSFs in the final stacks of the 
two sets look similar to the one shown in Fig. \ref{SGP_aniso}, since the
individual anisotropies in defocused exposures rotate by 90 degrees when the
detector plane passes through the focal plane, thus they average out in the
coaddition. From these two entirely independent data sets we confirm the shear
detections of EIS0045$-$2923 and LIL004608$-$292341, but with a lower $S/N$ 
due to the reduced number density ($n=19$ arcmin$^{-2}$) of galaxies. The
fluctuations of less significant peaks ($<3-3.5\sigma$) between
the various realisations increase, especially when moving to filter scales
below 10\myarcminnodot.

\section{Conclusions}
We presented a sample of shear-selected mass concentrations in a deep
observation of the South Galactic Pole, which was previously covered by the
EIS Wide Survey Patch B field. Two of the mass concentrations found coincide 
with the colour-selected cluster candidates EIS0045$-$2923 and
LIL004608$-$292341, taken from the Patch B observations, and thus confirm
their cluster nature. There is evidence for a possible filament connecting
LIL004608$-$292341 with LIL004636$-$293539. A smaller mass concentration found
is probably associated with EIS0045$-$2948. For the remaining low-$S/N$
detections no conclusions could be reached. We have shown that the
shear-selection method yields very useful results also for less massive
clusters.

\begin{acknowledgements}
This work was supported by the BMBF through the DLR under the project 50 OR 
0106, by the BMBF through DESY under the project 05AE2PDA/8, and by the 
Deutsche Forschungsgemeinschaft (DFG) under the project SCHN 342/3--1.
\end{acknowledgements}

\end{document}